\documentclass[conference]{IEEEtran}
\IEEEoverridecommandlockouts
\usepackage{cite}
\usepackage{amsmath,amssymb,amsfonts}
\usepackage{algorithmic}
\usepackage{graphicx}
\usepackage{textcomp}
\usepackage{xcolor}
\usepackage{booktabs}
\usepackage{multirow}
\def\BibTeX{{\rm B\kern-.05em{\sc i\kern-.025em b}\kern-.08em
    T\kern-.1667em\lower.7ex\hbox{E}\kern-.125emX}}
\begin{document}

\title{A Benchmark Study of Deep Learning Methods for Multi-Label Pediatric Electrocardiogram-Based Cardiovascular Disease Classification
}

\author{
\IEEEauthorblockN{Yiqiao Chen}
\IEEEauthorblockA{\textit{Department of Computer Science and Technology} \\
\textit{Beijing Institute of Technology, Zhuhai}\\
Zhuhai, China \\
Email: cyq2161@gmail.com}
}

\maketitle

\begin{abstract}
Cardiovascular disease (CVD) is a major pediatric health burden, and early screening is of critical importance. Electrocardiography (ECG), as a noninvasive and accessible tool, is well suited for this purpose. This paper presents the first benchmark study of deep learning for multi-label pediatric CVD classification on the recently released ZZU-pECG dataset, comprising 3716 recordings with 19 CVD categories. We systematically evaluate four representative paradigms—ResNet-1D, BiLSTM, Transformer, and Mamba 2—under both 9-lead and 12-lead configurations. All models achieved strong results, with Hamming Loss as low as 0.0069 and F1-scores above 85\% in most settings. ResNet-1D reached a macro-F1 of 94.67\% on the 12-lead subset, while BiLSTM and Transformer also showed competitive performance. Per-class analysis indicated challenges for rare conditions such as hypertrophic cardiomyopathy in the 9-lead subset, reflecting the effect of limited positive samples. This benchmark establishes reusable baselines and highlights complementary strengths across paradigms. It further points to the need for larger-scale, multi-center validation, age-stratified analysis, and broader disease coverage to support real-world pediatric ECG applications.
\end{abstract}


\section{Introduction}
\begin{figure*}[t]
	\centering
	\includegraphics[width=0.95\textwidth]{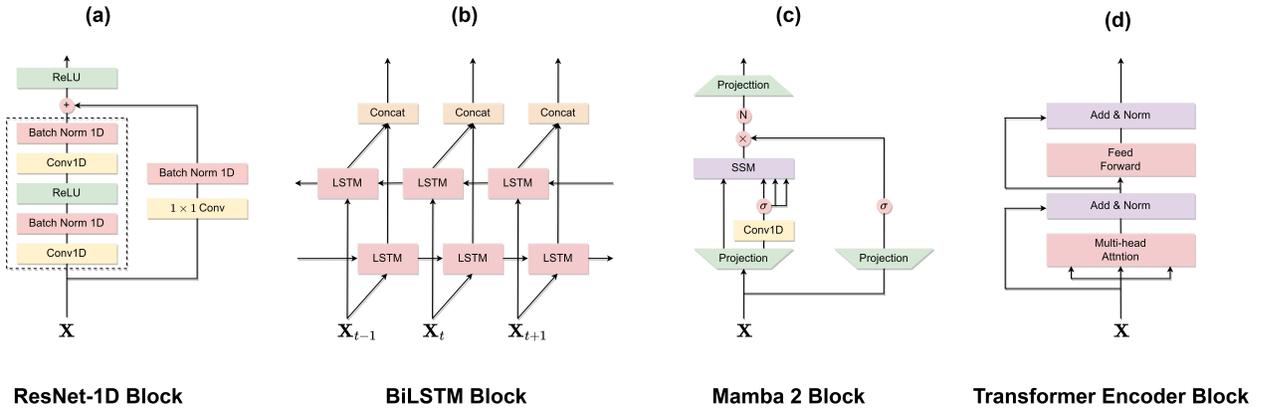}
	\caption{Schematic comparison of the four benchmark architectures evaluated in this study. From left to right: (a) a residual convolutional block (ResNet-1D), which captures local temporal patterns via stacked 1D convolutions and shortcut connections; (b) a bidirectional long short-term memory (BiLSTM) network, which models sequential dependencies in both forward and backward directions; (c) a Mamba 2 block, a recent state-space model (SSM) that combines input projection, depthwise convolution, and structured state-space dynamics for efficient long-range dependency modeling; (d) a Transformer encoder block, which leverages multi-head self-attention and feed-forward layers for global context modeling.}
	\label{schematic}
\end{figure*}
\par Cardiovascular disease (CVD) remains a leading cause of death worldwide\cite{roth2025global}. Cardiovascular risk factors and pathological changes often emerge during childhood and adolescence, and without early detection and intervention, they may accumulate into substantial burdens in adulthood \cite{kartiosuo2024cardiovascular}. Early risk assessment and screening in pediatric populations are therefore of critical public-health and clinical importance.
Electrocardiography (ECG) is a routine noninvasive test with low cost, high accessibility, and bedside convenience, making it well suited for large-scale pediatric screening and follow-up\cite{mayourian2024pediatric, baldazzi2026wavelet, chen2024congenital}.
In recent years, deep learning has achieved remarkable progress in adult ECG interpretation and abnormality detection\cite{ding2025advances}. However, corresponding studies in pediatric populations remain scarce\cite{leone2024artificial}. Pediatric ECGs exhibit pronounced age-dependent variations in rate, intervals, morphology, and amplitude, and their disease spectrum differs substantially from adults; consequently, methods developed for adult ECGs cannot be directly transferred\cite{bratincsak2020electrocardiogram, rustwick2014comparison}. These differences underscore the need for dedicated studies focusing specifically on pediatric ECGs.

\par Most publicly available ECG datasets focus on adults and often provide only electrocardiographic statements without disease-level (ICD-10) labels\cite{moody2001impact, wagner2020ptb}. Most existing studies on multi-label cardiovascular disease detection from ECG signals have relied on the MIMIC-IV\cite{johnson2023mimic} and its various extended versions that provide disease-level annotations\cite{strodthoff2024prospects, oladipo2025benchmarking}. Pediatric-specific datasets are even more scarce and typically limited in scale, restricting their usefulness for systematic deep learning investigations\cite{leone2024artificial, mayourian2024pediatric}. Furthermore, most available pediatric studies have concentrated on arrhythmias or single disease categories, with little progress toward multi-label diagnosis or comprehensive baseline evaluations spanning the full spectrum of cardiovascular disorders. The recently released ZZU-pECG dataset\cite{tan2025pediatric} addresses this gap by providing the largest open-source pediatric ECG collection to date, comprising over 14,000 recordings from more than 11,000 children. Distinctly, it includes patient-level diagnostic information encoded according to the World Health Organization’s International Classification of Diseases, Tenth Revision (ICD-10). To the best of our knowledge, no prior work has applied deep learning techniques to this dataset, making our study the first to systematically investigate its potential for multi-label pediatric CVD classification.
Against this backdrop, a benchmark study is needed to address the following: in pediatric ECG multi-label CVD classification based on ZZU-pECG dataset, which mainstream deep learning paradigms (CNNs, RNNs, Transformers, Mamba) are better suited, how large are the performance gaps, and what underlying factors (e.g., temporal, global, or long-sequence modeling) drive these differences. Such results would provide a reusable reference and practical boundaries for future model and task design.

\par In this study, we employ the recently released ZZU-pECG dataset and specifically analyze the 3,716 recordings annotated with cardiovascular disease labels.
We formulate the task as multi-label cardiovascular disease classification with 19 pediatric CVD labels, encompassing myocarditis, cardiomyopathies, several congenital heart disease subtypes, and Kawasaki disease.
For systematic comparison, we evaluate four representative deep learning models: ResNet-1D\cite{he2016deep}, BiLSTM\cite{hochreiter1997long}, Transformer\cite{vaswani2017attention}, and Mamba 2\cite{dao2024transformers}. All models are trained and tested consistently on both 9-lead and 12-lead subsets to assess the strengths and complementarity of temporal and global modeling paradigms in pediatric ECG multi-label diagnosis. In addition, we recognize and mitigate the issue of class imbalance inherent in this dataset, ensuring fairer evaluation across rare and common conditions.

\par In this work, we make the following key contributions:
\begin{enumerate}
    \item To the best of our knowledge, we are the first to apply deep learning methods to pediatric ECG data based on ZZU-pECG dataset for multi-label cardiovascular disease classification.
    \item Under a unified dataset and evaluation protocol, we compare four mainstream deep learning paradigms on both 9-lead and 12-lead subsets, establishing reusable baselines and upper-bound estimates to serve as direct references for future studies.
\end{enumerate}

\par The remainder of this paper is organized as follows: Section~\ref{Related Work} reviews prior deep learning studies on adult and pediatric ECGs; Section~\ref{Methods} describes the dataset and task formulation; Section~\ref{Results and Discussions} reports and analyzes the experimental results; and Section~\ref{Conclusion} concludes the paper and outlines directions for future work.

\section{Related Work}\label{Related Work}
\par ECG-based deep learning methods have been extensively applied to both the diagnosis and prognosis of cardiovascular diseases, spanning tasks such as arrhythmia detection, myocardial infarction recognition, heart failure classification, and risk stratification for adverse outcomes \cite{moreno2024ecg}. For instance, an automated system for multi-class classification of normal rhythm, coronary artery disease, myocardial infarction, and congestive heart failure achieved over 98.5\% accuracy using a novel GaborCNN\cite{jahmunah2021automated}. Similarly, a lightweight hybrid CNN–LSTM model has been proposed for arrhythmia detection, delivering high accuracy and offering interpretability through explainable analysis of ECG features \cite{alamatsaz2024lightweight}. More recently, a hierarchical multi-stream deep learning framework incorporating similarity maps has been introduced for ventricular arrhythmia classification, showing robust performance across multiple public ECG datasets and strong generalization to unseen data \cite{lin2024ventricular}. Beyond these task-specific classification efforts, an increasingly prominent line of research is disease-level multi-label classification of cardiovascular conditions. For example, a recent benchmark study systematically evaluated six state-of-the-art deep learning architectures on the MIMIC-IV-ECG dataset, comparing CNN, CNN–BiLSTM, spectrogram-based CNN, hierarchical attention network, CAT-Net, and S4 models for multi-label heart disease prediction under standardized protocols \cite{oladipo2025benchmarking}.
\par However, most of the aforementioned studies have been conducted on adult cohorts, while pediatric-specific investigations remain limited. 
One line of work developed and externally validated a convolutional neural network trained on nearly 100,000 ECG–echocardiogram pairs to predict left ventricular dysfunction and remodeling in children, while also providing interpretability through saliency mapping \cite{mayourian2024pediatric}. Another effort introduced a few-shot learning framework that leverages multimodal learning and a novel class-aware contrastive loss, termed AGCACL, which incorporates hard mining strategies to enhance pediatric arrhythmia classification under limited data conditions \cite{chen2025advancing}. Yet, to the best of our knowledge, no prior work has systematically addressed disease-level multi-label classification of pediatric cardiovascular diseases. Overall, despite extensive progress in adult ECG research and emerging pediatric efforts, the absence of systematic benchmarks for disease-level multi-label classification in children highlights a critical gap. Addressing this gap forms the central motivation of our work.

\section{Methods}\label{Methods}
\subsection{Dataset}
\begin{table}[htbp]
\centering
\renewcommand{\arraystretch}{1.3} 
\caption{List of 19 cardiovascular disease (CVD) labels and their ICD-10 codes used in this study}
\label{class}
\resizebox{0.48\textwidth}{!}{   
\begin{tabular}{lll}
\toprule
Label & Disease & ICD-10 \\
\midrule
1  & Fulminant myocarditis & (F) I40.0 \\
2  & Viral myocarditis & (V) I40.0 \\
3  & Acute myocarditis & I40.9 \\
4  & Myocarditis & I51.4 \\
5  & Dilated cardiomyopathy & I42.0 \\
6  & Hypertrophic cardiomyopathy & I42.2 \\
7  & Cardiomyopathy & I42.9 \\
8  & Noncompaction of ventricular myocardium & Q24.8 \\
9  & Kawasaki disease & M30.3 \\
10 & Ventricular septal defect & Q21.0 \\
11 & Atrial septal defect & Q21.1 \\
12 & ASD (Foramen ovale) & (FO) Q21.1 \\
13 & ASD (Ostium secundum defect) & (OSD) Q21.1 \\
14 & Atrioventricular septal defect & Q21.2 \\
15 & Tetralogy of Fallot & Q21.3 \\
16 & Stenosis of RV outflow tract & Q22.1 \\
17 & Patent ductus arteriosus & Q25.0 \\
18 & Stenosis of pulmonary artery & Q25.6 \\
19 & Pulmonary valve stenosis & I37.0 \\
\bottomrule
\end{tabular}
}
\end{table}

\begin{table}[htbp]
	\centering
	\renewcommand{\arraystretch}{1.2}
	\caption{Hamming Loss comparison of models on 9-lead and 12-lead ECG classification}
	\label{tab:hamming_loss}
	\resizebox{0.40\textwidth}{!}{
		\begin{tabular}{lcc}
			\toprule
			Methods & 9-lead & 12-lead \\
			\midrule
			ResNet-1D   & \textbf{0.0076} & \textbf{0.0069} \\
			BiLSTM      & 0.0148 & 0.0092 \\
			Mamba 2     & 0.0124 & 0.0188 \\
			Transformer & 0.0178 & 0.0116 \\
			\bottomrule
		\end{tabular}
	}
\end{table}

\begin{table*}[htbp]
	\centering
	\renewcommand{\arraystretch}{1.2}
	\caption{Performance comparison of deep learning models on 9-lead ECG multi-label classification}
	\label{tab:results_9lead}
	\resizebox{\textwidth}{!}{
		\begin{tabular}{lcccccccc}
			\toprule
			\multirow{2}{*}{Methods} & \multicolumn{4}{c}{Micro (\%)} & \multicolumn{4}{c}{Macro (\%)} \\
			\cmidrule(lr){2-5} \cmidrule(lr){6-9}
			& Precision & Recall & F1 & AUC-ROC & Precision & Recall & F1 & AUC-ROC \\
			\midrule
			ResNet-1D   & \textbf{94.27} & \textbf{94.87} & \textbf{94.57} & \textbf{99.80} & \textbf{91.75} & \textbf{87.96} & \textbf{89.15} & 99.21 \\
			BiLSTM      & 88.37 & 90.80 & 89.57 & 99.10 & 82.84 & 80.58 & 79.75 & 98.27 \\
			Mamba 2     & 90.93 & 91.35 & 91.14 & 99.50 & 88.15 & 84.70 & 85.63 & \textbf{99.22} \\
			Transformer & 86.24 & 88.72 & 87.46 & 99.16 & 84.59 & 80.87 & 81.82 & 97.84 \\
			\bottomrule
		\end{tabular}
	}
\end{table*}

\begin{table*}[htbp]
	\centering
	\renewcommand{\arraystretch}{1.2}
	\caption{Performance comparison of deep learning models on 12-lead ECG multi-label classification}
	\label{tab:results_12lead}
	\resizebox{\textwidth}{!}{
		\begin{tabular}{lcccccccc}
			\toprule
			\multirow{2}{*}{Methods} & \multicolumn{4}{c}{Micro (\%)} & \multicolumn{4}{c}{Macro (\%)} \\
			\cmidrule(lr){2-5} \cmidrule(lr){6-9}
			& Precision & Recall & F1 & AUC-ROC & Precision & Recall & F1 & AUC-ROC \\
			\midrule
			ResNet-1D   & \textbf{93.62} & 94.53 & \textbf{94.07} & \textbf{99.80} & \textbf{94.93} & \textbf{94.52} & \textbf{94.67} & \textbf{99.74} \\
			BiLSTM      & 89.63 & \textbf{95.24} & 92.35 & 99.58 & 89.17 & 93.25 & 90.96 & 99.38 \\
			Mamba 2     & 80.24 & 89.74 & 84.73 & 99.03 & 80.17 & 89.15 & 84.30 & 98.91 \\
			Transformer & 88.17 & 92.52 & 90.29 & 99.48 & 89.79 & 90.87 & 90.08 & 99.20 \\
			\bottomrule
		\end{tabular}
	}
\end{table*}
\par In this study, we employ the ZZU pECG pediatric electrocardiogram dataset, which spans hospitalized children aged 0–14 years from 2018 to 2024. The dataset comprises 11,643 subjects and 14,190 ECG recordings (12,334 with 12 leads and 1,856 with 9 leads), all sampled uniformly at 500 Hz with recording durations between 5 and 120 seconds. In addition to conventional ECG diagnostic statements, the dataset provides disease diagnoses with ICD-10 codes, and all original diagnostic statements have been mapped to standardized codes following both the AHA and CHN guidelines, enabling cross-study comparability and multi-task modeling. Unlike prior repositories that predominantly focus on adult populations and lack explicit disease annotations, ZZU pECG explicitly covers 19 categories of pediatric cardiovascular diseases (e.g., myocarditis, cardiomyopathy, congenital heart disease, Kawasaki disease). This makes it particularly suitable for multi-label classification and disease recognition benchmarks in pediatric ECG analysis.
\par First, we extracted 3,716 ECG records containing cardiovascular disease diagnoses and divided them into two subsets according to lead configuration: 12-lead and 9-lead. This separation is motivated by a clinical factor in the dataset: the number of leads is partially associated with age, where 9-lead ECGs are primarily recorded in younger children (under the age of 7) due to the difficulty of placing all six chest leads.
\par In preliminary experiments with a strict subject-wise split, we observed that certain disease classes had positive samples restricted to a single partition (training, validation, or test), which led to missing evaluation metrics or distorted results. To ensure feasible and reliable benchmarking, we therefore adopted a slice-then-split strategy: each original recording was segmented into fixed 3-second windows. To reduce computational cost, all samples were uniformly down-sampled by a factor of 5. Finally, we performed a stratified random split with a ratio of 7:1:2 for training, validation, and testing dataset.
\par This approach yields a more balanced distribution of classes across partitions, mitigating the adverse effects of extreme class imbalance—an issue particularly critical in clinical datasets. The final benchmark task covers 19 disease categories, as shown in Table \ref{class}. Notably, all samples in the 9-lead subset are free of the label 7 (Cardiomyopathy).
\begin{figure}[htbp]
    \centering
    \includegraphics[width=0.5\textwidth]{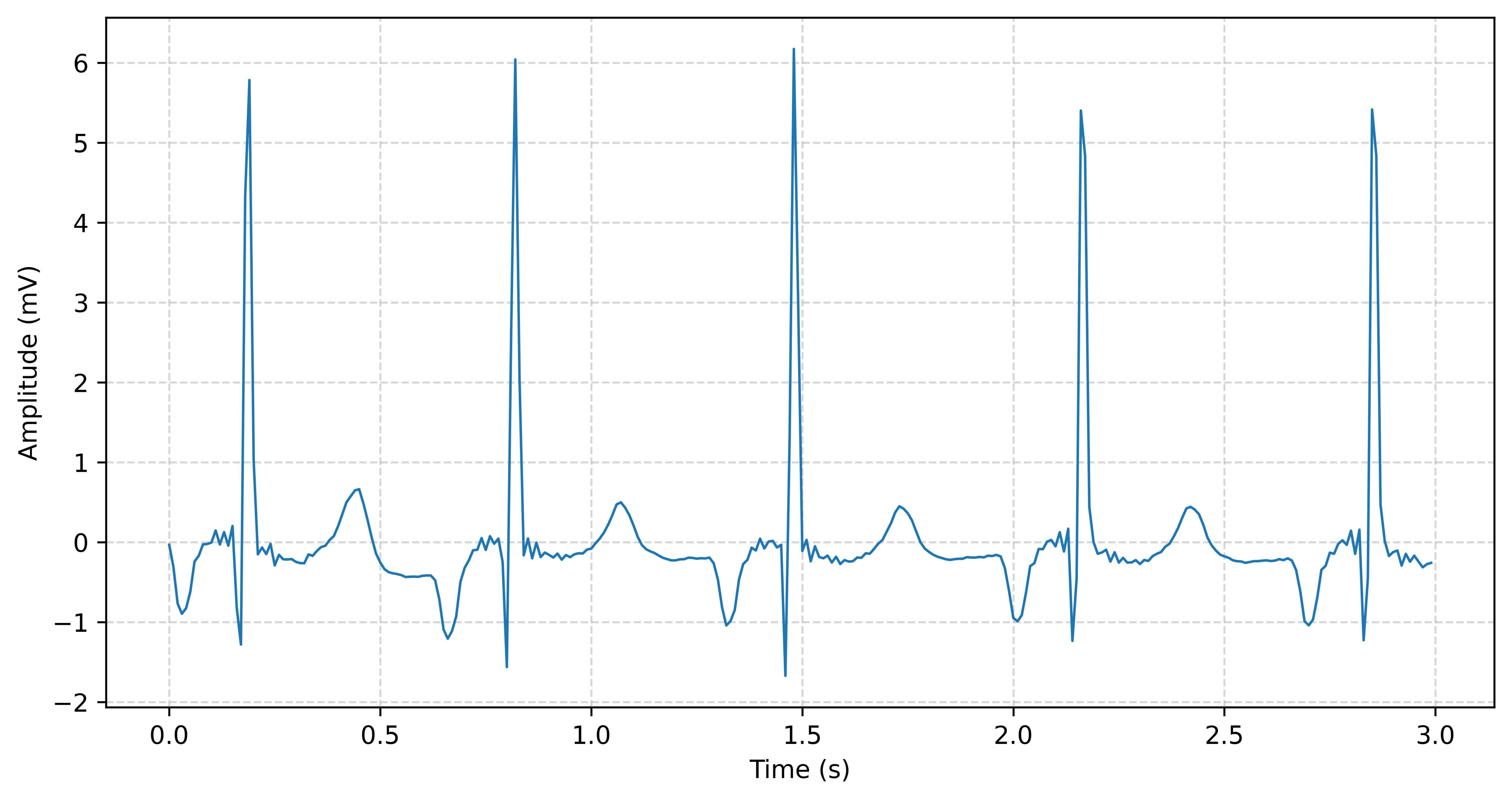}
    \caption{Example of a single-lead ECG slice.}
    \label{ecg_slice}
\end{figure}
\par Due to the dataset authors emphasized in the original publication that the ECG signals were of sufficiently high quality. Therefore, in this work we do not apply any additional denoising procedures; instead, we only perform per-lead normalization independently within the training, validation, and test sets. As shown in Fig.\ref{ecg_slice}, we illustrate a representative 3-second slice from a single lead.

\subsection{Baseline Models}
\par Given the absence of prior systematic baselines on pediatric ECGs, we establish reference points by training and evaluating four mainstream architectures: ResNet-1D, BiLSTM, Transformer, and Mamba 2, as illustrated in Fig.~\ref{schematic}. These models span different design philosophies—convolutional, recurrent, attention-based, and state-space—and thus provide a comprehensive view of model behavior on this task.
\par For the convolutional neural network baseline, we adopt a one-dimensional residual network (ResNet-1D). This architecture employs residual connections to alleviate gradient vanishing and support deeper feature extraction. The model consists of four residual stages, where each basic block (BasicBlock1D) contains two one-dimensional convolutional layers (kernel size = 7), batch normalization, and ReLU activation, with an optional downsampling branch to ensure dimension matching. As the network deepens, the channel width is progressively expanded to 64, 128, 256, and 512, while stride-2 convolutions and pooling operations gradually reduce sequence length. After feature extraction, a global adaptive average pooling layer aggregates the temporal dimension into a fixed-length 512-dimensional vector, which is subsequently fed into a fully connected layer that outputs predictions across 19 disease categories. Notably, for ECG inputs with 9 leads and 12 leads, the number of input channels for the first convolutional layer is set to 9 and 12, respectively, ensuring compatibility with different lead configurations.
\par For the recurrent neural network baseline, we adopt a one-dimensional bidirectional long short-term memory network (BiLSTM). Unlike convolutional networks, LSTMs emphasize sequential modeling and can capture the temporal dependencies of ECG signals. In our design, the input feature dimension at each time step corresponds to the number of ECG leads (9 or 12). The network consists of two stacked bidirectional LSTM layers, each with a hidden size of 128, resulting in a concatenated output dimension of 256. To mitigate overfitting, a dropout rate of 0.5 is applied between LSTM layers. Along the sequence dimension, we retain only the hidden state of the last time step as the global representation, which is then fed into a fully connected layer to predict across 19 cardiovascular disease categories.

\begin{figure*}[htbp]
	\centering
	\includegraphics[width=0.95\textwidth]{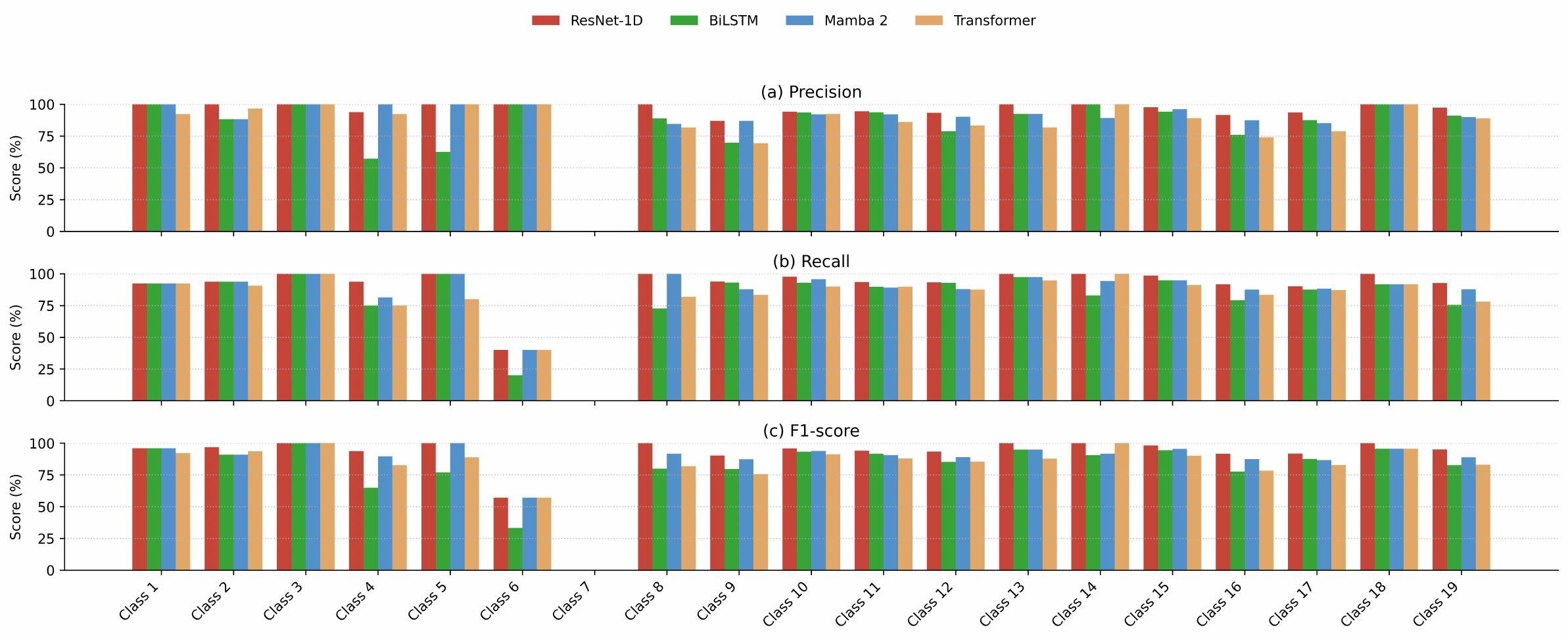}
	\caption{Per-class performance comparison of four baseline models (ResNet-1D, BiLSTM, Mamba 2, and Transformer) on the 9-lead ECG subset. Panels (a), (b), and (c) respectively show the precision, recall, and F1-score for each of the 19 diagnostic classes. Scores are reported in percentage. Note that Class 7 was excluded from evaluation due to the absence of positive samples in this subset.}
	\label{9 lead}
\end{figure*}

\begin{figure*}[htbp]
	\centering
	\includegraphics[width=0.95\textwidth]{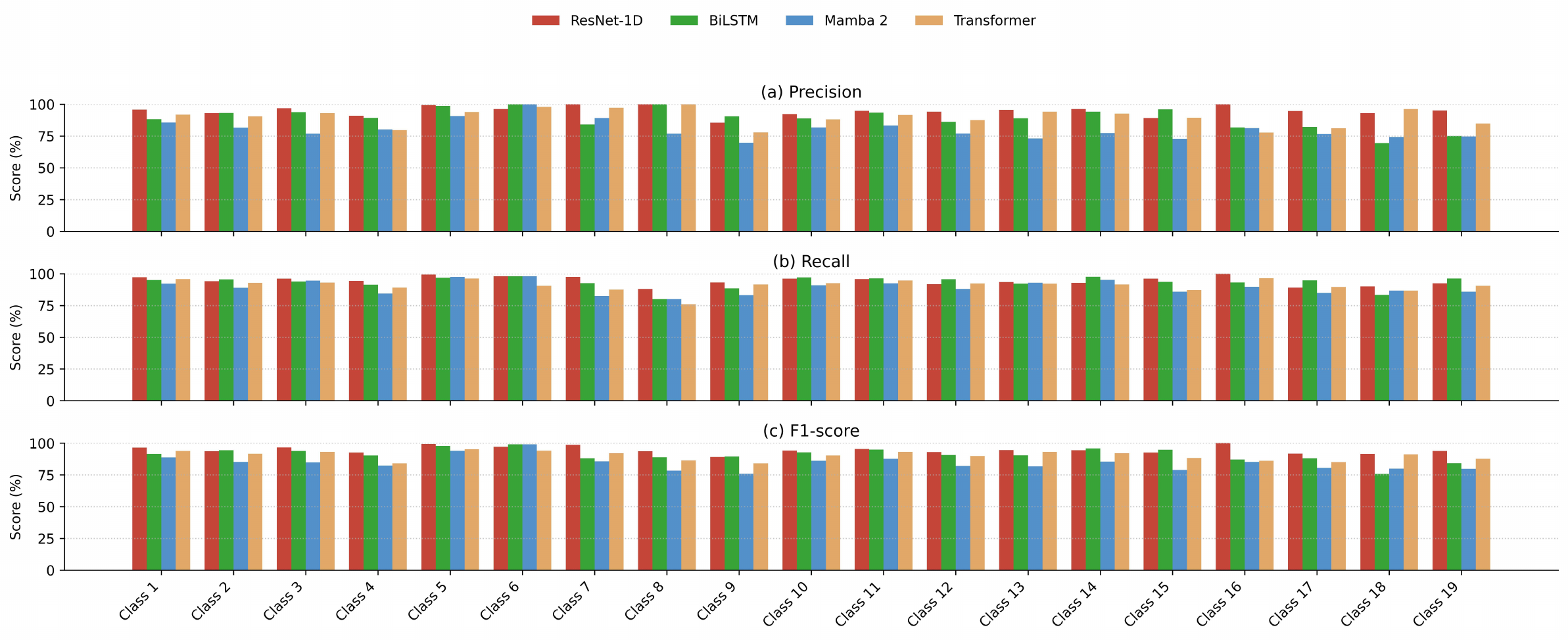}
	\caption{Per-class performance comparison of four baseline models (ResNet-1D, BiLSTM, Mamba 2, and Transformer) on the 12-lead ECG subset. Panels (a), (b), and (c) respectively show the precision, recall, and F1-score for each of the 19 diagnostic classes. Scores are reported in percentage.}
	\label{12 lead}
\end{figure*}

\begin{figure*}[t]
	\centering
	\includegraphics[width=0.95\textwidth]{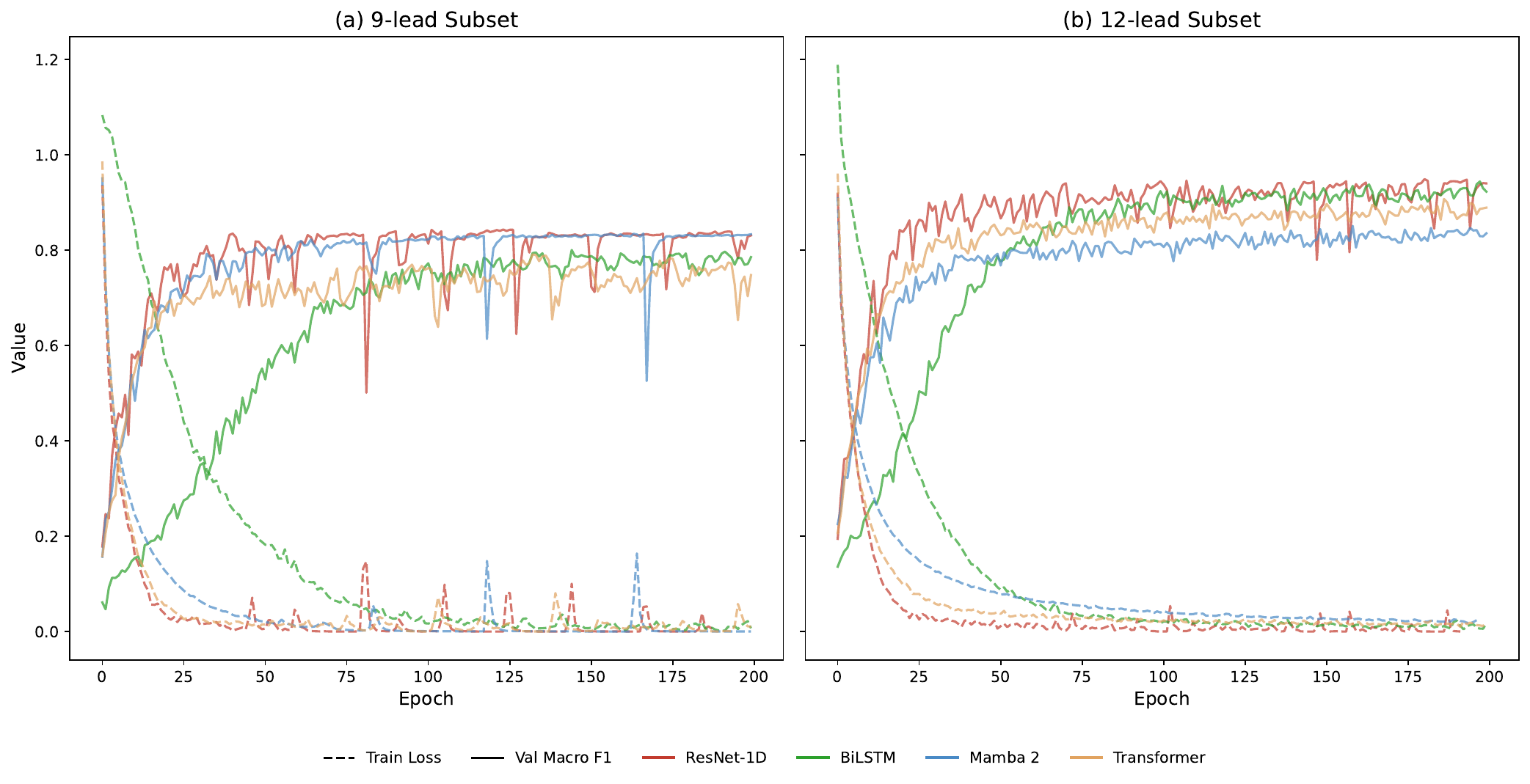}
	\caption{Training dynamics of four baseline models (ResNet-1D, BiLSTM, Mamba 2, and Transformer) on the pediatric ECG multi-label CVD classification task. Panels (a) and (b) show results on the 9-lead and 12-lead subsets, respectively. Solid lines denote macro F1 scores on validation dataset.}
	\label{loss f1}
\end{figure*}

\par For the state-space model baseline, we adopt a Mamba 2 based classifier. The model first applies a linear projection to map the raw ECG input channels (9 or 12 leads) into a latent embedding space of dimension 128, which serves as the effective feature dimension throughout the network. The backbone comprises two stacked Mamba 2 layers configured with a state dimension of 64, a convolutional kernel size of 4, an expansion factor of 2, a per-head hidden size of 8, and a chunk size of 30, enabling efficient sequence partitioning for the 300-sample input sequences. Following sequence modeling, the hidden representations are transposed and aggregated using global adaptive average pooling to obtain a fixed-length vector of dimension 128 for each ECG recording. A dropout rate of 0.2 is applied during training to improve generalization, and the final fully connected classification head outputs predictions across 19 disease categories. This design leverages the structured state-space formulation of Mamba 2 to capture long-range dependencies in a computationally efficient manner, providing a strong complementary baseline to convolutional, recurrent, and attention-based approaches.
\par For the attention-based baseline, we adopt a Transformer encoder classifier. Similar to the Mamba 2 model, the raw ECG inputs with 9 or 12 leads are first transformed into a latent embedding space of dimension 128, which serves as the effective feature dimension throughout the network. To achieve this, we apply a patch embedding strategy: the input sequence of length 300 is divided into non-overlapping segments of size 3 along the temporal dimension, yielding 100 patches. For each patch, the signals across all leads are flattened into a vector and linearly projected into the 128-dimensional embedding space. This operation reduces the effective sequence length while preserving local temporal structures, making the subsequent attention mechanism more efficient. Positional information is then incorporated by adding fixed sinusoidal encodings to the embedded sequence. The sequence is subsequently processed by two stacked Transformer decoder layers, each consisting of multi-head self-attention (head dimension is 8) followed by a feedforward network with hidden size 256 and dropout rate 0.1. After temporal modeling, the hidden representations are aggregated using global adaptive average pooling to obtain a fixed-length 128-dimensional vector. This vector is further passed through a classification head composed of normalization and multiple fully connected layers, producing predictions over 19 cardiovascular disease categories. By adopting the same initial projection strategy as Mamba 2 and explicitly incorporating patch-based embedding, this design ensures a fairer comparison across baselines while highlighting the complementary advantages of attention-based sequence modeling for ECG classification. It is worth noting that among the four baselines, only the Transformer model adopts patch embedding. This design choice follows the standard practice of attention-based sequence modeling, where patching reduces the effective sequence length and mitigates the quadratic computational cost of self-attention while retaining local temporal structures. In contrast, recurrent networks such as BiLSTM are naturally designed to process sequences in a step-by-step manner, making direct pointwise input more faithful to their sequential inductive bias. Similarly, Mamba 2 is specifically optimized for efficient long-sequence modeling through state space formulations and thus does not require patching. Preserving these canonical input strategies across paradigms ensures both fairness in comparison and faithfulness to each model’s intrinsic design philosophy.
\par In summary, our benchmark covers four representative paradigms—convolutional, recurrent, attention-based, and state-space—each evaluated under its canonical design. This setup allows us to highlight their complementary modeling advantages: CNNs excel at local feature extraction, BiLSTMs capture sequential dependencies, Transformers leverage patching and attention for global context, and Mamba 2 efficiently handles long-range dynamics. By preserving the natural input strategies of each model, we ensure both fairness and interpretability in cross-paradigm comparisons. It is important to note that in the 9-lead subset, the seventh disease category (Cardiomyopathy) does not appear in the test data at all. Therefore, although the classification head of our model still produces 19-dimensional predictions, this category is skipped during the evaluation of performance metrics in order to avoid invalid or misleading results. We deliberately retain the 19-dimensional output design for two reasons. First, this ensures consistency with the 12-lead models, allowing fair comparison across different lead configurations. Second, although Cardiomyopathy does not occur in the current 9-lead subset, this is most likely a consequence of the dataset’s scale and distribution rather than a true absence in practice; in real-world clinical scenarios, such cases do exist. For this reason, in the per-class results reported later for the 9-lead subset, the seventh category is not included, while the overall task is still formulated as a 19-class classification problem.

\subsection{Experimental Setup and Training Strategies}
In this study, all experiments were conducted on a workstation equipped with an AMD Ryzen 7 7435H CPU, 32 GB RAM, and a single NVIDIA GeForce RTX 4070 Laptop GPU with 8 GB memory. The operating system was Windows 11 (64-bit), and all models were implemented in PyTorch. All models were trained for 200 epochs with a batch size of 64 using the AdamW\cite{loshchilov2017decoupled} optimizer. The initial learning rate was fixed at 7e-4 without weight decay or learning rate scheduling. At inference time, to ensure fairness across baselines, the decision threshold is uniformly fixed at 0.5.
\par Here, we adopt a weighted binary cross-entropy loss with logits to account for the severe class imbalance. 
Given a sample $x_i$, the model outputs logits $\mathbf{z}_i \in \mathbb{R}^C$ with ground-truth multi-hot labels $\mathbf{y}_i \in \{0,1\}^C$. 
The per-class loss for $i$-th sample in a mini-batch is defined as:
\begin{equation}
\mathcal{L}_{\text{BCE}}^{i,c} 
= (1-y_{i,c}) \cdot \log\!\big(1+e^{z_{i,c}}\big) 
+ y_{i,c} \cdot w_c \cdot \log\!\big(1+e^{-z_{i,c}}\big)
\label{eq:loss}
\end{equation}
where $z_{i,c}$ denotes the logit predicted for class $c$, $y_{i,c}\in\{0,1\}$ is the ground-truth label, and $w_c$ is the positive class weight, which balances the contributions of positive and negative samples:
\begin{equation}
w_c = \min\!\left( \frac{N - \sum_{i=1}^N y_{i,c}}{\max\!\big(1, \sum_{i=1}^N y_{i,c}\big)}, \; \tau \right)
\label{eq:posweight}
\end{equation}
with $N$ the number of training instances and $\tau$ a clipping threshold (set to $100$ in our experiments). 
This weighting increases the influence of minority-class positives, thereby alleviating the imbalance problem inherent in multi-label classification.

\section{Results and Discussions}\label{Results and Discussions}
\par As shown in Table \ref{tab:hamming_loss}, Table \ref{tab:results_9lead}, and Table \ref{tab:results_12lead}, all models achieved near-ideal performance on the task. For instance, ResNet-1D reached a Hamming Loss of only 0.0069 and a macro-averaged F1-score of 94.67\% on the 12-lead subset, while most other models also maintained F1-scores above 85\%. These results highlight the strong potential of deep learning approaches for multi-label pediatric ECG classification. Nevertheless, it should be noted that the evaluation may not fully guarantee subject-level independence, and thus the reported numbers may partially reflect optimistic estimates. Across paradigms, convolutional architectures demonstrated the most stable and prominent performance, underscoring the discriminative power of local morphological features and establishing ResNet-1D as both a reliable baseline and an upper-bound reference. BiLSTM exhibited slightly higher micro-recall, confirming the contribution of temporal dependency modeling in certain categories. Transformer showed substantial improvements in the 12-lead setting, suggesting that its capacity for global context modeling benefits more fully from larger sample sizes. In contrast, Mamba 2 obtained a markedly lower precision (80.24\%) in the 12-lead condition, possibly due to suboptimal hyperparameter settings rather than intrinsic limitations of the paradigm. Overall, the four paradigms present complementary strengths and weaknesses, offering guidance for model selection in practical clinical scenarios depending on data availability and task requirements.

\par The per-class results, illustrated in Fig. \ref{9 lead} and Fig. \ref{12 lead}, further reveal the heterogeneity of model behavior across diagnostic categories. While most classes achieved high precision and recall under both lead configurations, the 9-lead subset showed noticeable difficulty in class 6 (Hypertrophic Cardiomyopathy), where performance was substantially lower across all paradigms. This observation is likely influenced by the limited number of positive cases for this condition in the 9-lead data, highlighting the sensitivity of rare categories to sampling constraints rather than reflecting systematic shortcomings of specific models.

The training dynamics, illustrated in Fig. \ref{loss f1}, highlight both convergence behaviors and generalization stability across paradigms. On the 9-lead subset, all models converged within the first 50–100 epochs, though with varying speeds and levels of fluctuation. ResNet-1D rapidly reached a stable plateau with relatively smooth F1 trajectories, whereas BiLSTM and Transformer required more epochs to approach comparable performance and displayed more pronounced variance during training. Mamba 2 exhibited a gradual rise before stabilization, reflecting its distinct optimization dynamics. On the 12-lead subset, convergence was generally faster and validation F1 scores attained higher and more stable levels across all models. This suggests that the increased sample size and richer input information facilitate more robust generalization. Importantly, while absolute scores differed, all paradigms eventually demonstrated consistent convergence without signs of severe overfitting, underscoring the feasibility of applying diverse deep architectures under both lead configurations.

\par A comparison between the 9-lead and 12-lead subsets provides additional insights into dataset-related factors. In the 9-lead test set, class 7 (Cardiomyopathy) contained no positive cases, which objectively increases the variability of some aggregate metrics. Moreover, the 9-lead subset is notably smaller in size compared with the 12-lead subset, making it difficult to draw firm conclusions about the effect of lead configuration on model performance. Further assessment of this question will require larger-scale data or external validation across institutions. From a practical perspective, however, the results suggest potential considerations for deployment. In scenarios where only a reduced number of leads can be collected, one may prefer paradigms that demonstrated relatively stable performance under the 9-lead setting (e.g. ResNet-1D, Mamba 2). In contrast, when the full 12-lead configuration is available, model selection can be made more flexibly, with the possibility of balancing precision and recall according to clinical priorities such as screening versus triage.

\par This benchmark highlights several directions for future research and clinical translation. First, larger-scale studies are needed to strengthen the evidence base, particularly through multi-center external validation and stratified analyses across different pediatric age groups. Moreover, while this work focused exclusively on cardiovascular disease (CVD) labels, future efforts could extend to broader disease coverage beyond CVD, enabling more comprehensive clinical utility. From a methodological perspective, the relatively stable performance of ResNet-1D across both lead configurations suggests that local temporal morphological features may provide generalizable cues for multi-disease recognition. The more pronounced improvement of the Transformer in the 12-lead setting may reflect the benefits of larger sample sizes for global context modeling. These observations, while preliminary, motivate further investigation into how model inductive biases interact with data scale and input richness. Finally, a key limitation of this study lies in the dataset size and distribution, which constrained the ability to perform strictly subject-independent training and evaluation. Addressing this challenge will require not only larger and more diverse datasets, but also methodological advances that enhance cross-subject generalization in real-world clinical settings.

\section{Conclusion}\label{Conclusion}
In this study, we conducted the first systematic benchmark of deep learning methods for multi-label pediatric ECG-based cardiovascular disease classification using the ZZU-pECG dataset. By evaluating four representative paradigms—ResNet-1D, BiLSTM, Transformer, and Mamba 2—across both 9-lead and 12-lead subsets, we established performance baselines and identified complementary modeling behaviors. All models achieved high performance, with F1-scores generally above 85\% and Hamming Loss values below 0.02, underscoring the feasibility of applying diverse architectures to pediatric ECG analysis. Our results also revealed challenges inherent in rare categories and limited subsets, as seen in the 9-lead setting where certain classes lacked positive cases. These findings emphasize the importance of larger datasets, multi-center external validation, and stratified evaluations across age groups. Furthermore, the observed stability of convolutional models and the performance gains of attention-based models in larger sample settings suggest avenues for exploring the interaction between architectural inductive biases and data richness. Overall, this benchmark provides both reusable baselines and practical insights for future research on pediatric ECGs. Expanding disease coverage beyond CVD and advancing cross-subject generalization remain critical steps toward real-world clinical deployment. To mitigate the risk of cross-subject information leakage, future work will adopt strictly patient-level data splits, ensuring that all ECG segments from the same patient remain confined to a single dataset partition. We also plan to incorporate multi-center validation and advanced augmentation strategies to further enhance cross-subject generalization.

\section*{Acknowledgment}
We are deeply grateful to the patients and their families for their participation and support, and we sincerely hope that this research may contribute to improved clinical care in the future. We also extend our gratitude to the dataset creators for their commitment to open science and for making the data publicly available.

\bibliographystyle{IEEEtran}

\begin{thebibliography}{10}
	\providecommand{\url}[1]{#1}
	\csname url@samestyle\endcsname
	\providecommand{\newblock}{\relax}
	\providecommand{\bibinfo}[2]{#2}
	\providecommand{\BIBentrySTDinterwordspacing}{\spaceskip=0pt\relax}
	\providecommand{\BIBentryALTinterwordstretchfactor}{4}
	\providecommand{\BIBentryALTinterwordspacing}{\spaceskip=\fontdimen2\font plus
		\BIBentryALTinterwordstretchfactor\fontdimen3\font minus
		\fontdimen4\font\relax}
	\providecommand{\BIBforeignlanguage}[2]{{%
			\expandafter\ifx\csname l@#1\endcsname\relax
			\typeout{** WARNING: IEEEtran.bst: No hyphenation pattern has been}%
			\typeout{** loaded for the language `#1'. Using the pattern for}%
			\typeout{** the default language instead.}%
			\else
			\language=\csname l@#1\endcsname
			\fi
			#2}}
	\providecommand{\BIBdecl}{\relax}
	\BIBdecl
	
	\bibitem{roth2025global}
	G.~A. Roth, G.~Collaborators \emph{et~al.}, ``Global, regional, and national
	burden of cardiovascular diseases and risk factors in 204 countries and
	territories, 1990-2023,'' \emph{Available at SSRN 5392535}, 2025.
	
	\bibitem{kartiosuo2024cardiovascular}
	N.~Kartiosuo, O.~T. Raitakari, M.~Juonala, J.~S. Viikari, A.~R. Sinaiko, A.~J.
	Venn, D.~R. Jacobs, E.~M. Urbina, J.~G. Woo, J.~Steinberger \emph{et~al.},
	``Cardiovascular risk factors in childhood and adulthood and cardiovascular
	disease in middle age,'' \emph{JAMA Network Open}, vol.~7, no.~6, pp.
	e2\,418\,148--e2\,418\,148, 2024.
	
	\bibitem{mayourian2024pediatric}
	J.~Mayourian, W.~G. La~Cava, A.~Vaid, G.~N. Nadkarni, S.~J. Ghelani, R.~Mannix,
	T.~Geva, A.~Dionne, M.~E. Alexander, S.~Q. Duong \emph{et~al.}, ``Pediatric
	ecg-based deep learning to predict left ventricular dysfunction and
	remodeling,'' \emph{Circulation}, vol. 149, no.~12, pp. 917--931, 2024.
	
	\bibitem{baldazzi2026wavelet}
	G.~Baldazzi, M.~Corda, G.~Solinas, and D.~Pani, ``Wavelet-based algorithms for
	noninvasive fetal ecg post-processing: A methodological review,''
	\emph{Biomedical Signal Processing and Control}, vol. 111, p. 108350, 2026.
	
	\bibitem{chen2024congenital}
	J.~Chen, S.~Huang, Y.~Zhang, Q.~Chang, Y.~Zhang, D.~Li, J.~Qiu, L.~Hu, X.~Peng,
	Y.~Du \emph{et~al.}, ``Congenital heart disease detection by pediatric
	electrocardiogram based deep learning integrated with human concepts,''
	\emph{Nature Communications}, vol.~15, no.~1, p. 976, 2024.
	
	\bibitem{ding2025advances}
	C.~Ding, T.~Yao, C.~Wu, and J.~Ni, ``Advances in deep learning for personalized
	ecg diagnostics: A systematic review addressing inter-patient variability and
	generalization constraints,'' \emph{Biosensors and Bioelectronics}, vol. 271,
	p. 117073, 2025.
	
	\bibitem{leone2024artificial}
	D.~M. Leone, D.~O’Sullivan, and K.~Bravo-Jaimes, ``Artificial intelligence in
	pediatric electrocardiography: A comprehensive review,'' \emph{Children},
	vol.~12, no.~1, p.~25, 2024.
	
	\bibitem{bratincsak2020electrocardiogram}
	A.~Bratincs{\'a}k, C.~Kimata, B.~N. Limm-Chan, K.~P. Vincent, M.~R. Williams,
	and J.~C. Perry, ``Electrocardiogram standards for children and young adults
	using z-scores,'' \emph{Circulation: Arrhythmia and Electrophysiology},
	vol.~13, no.~8, p. e008253, 2020.
	
	\bibitem{rustwick2014comparison}
	B.~A. Rustwick and D.~L. Atkins, ``Comparison of electrocardiographic
	characteristics of adults and children for automated external defibrillator
	algorithms,'' \emph{Pediatric Emergency Care}, vol.~30, no.~12, pp. 851--855,
	2014.
	
	\bibitem{moody2001impact}
	G.~B. Moody and R.~G. Mark, ``The impact of the mit-bih arrhythmia database,''
	\emph{IEEE engineering in medicine and biology magazine}, vol.~20, no.~3, pp.
	45--50, 2001.
	
	\bibitem{wagner2020ptb}
	P.~Wagner, N.~Strodthoff, R.-D. Bousseljot, D.~Kreiseler, F.~I. Lunze,
	W.~Samek, and T.~Schaeffter, ``Ptb-xl, a large publicly available
	electrocardiography dataset,'' \emph{Scientific data}, vol.~7, no.~1, pp.
	1--15, 2020.
	
	\bibitem{johnson2023mimic}
	A.~E. Johnson, L.~Bulgarelli, L.~Shen, A.~Gayles, A.~Shammout, S.~Horng, T.~J.
	Pollard, S.~Hao, B.~Moody, B.~Gow \emph{et~al.}, ``Mimic-iv, a freely
	accessible electronic health record dataset,'' \emph{Scientific data},
	vol.~10, no.~1, p.~1, 2023.
	
	\bibitem{strodthoff2024prospects}
	N.~Strodthoff, J.~M. Lopez~Alcaraz, and W.~Haverkamp, ``Prospects for
	artificial intelligence-enhanced electrocardiogram as a unified screening
	tool for cardiac and non-cardiac conditions: an explorative study in
	emergency care,'' \emph{European Heart Journal-Digital Health}, vol.~5,
	no.~4, pp. 454--460, 2024.
	
	\bibitem{oladipo2025benchmarking}
	E.~Oladipo, S.~Nazrul, and M.~Nafea, ``Benchmarking deep learning architectures
	for ecg-based multi-label heart disease prediction using mimic-iv database,''
	in \emph{2025 IEEE 38th International Symposium on Computer-Based Medical
		Systems (CBMS)}.\hskip 1em plus 0.5em minus 0.4em\relax IEEE, 2025, pp.
	795--800.
	
	\bibitem{tan2025pediatric}
	J.~Tan, H.~Fan, J.~Luo, Y.~Zhou, N.~Wang, X.~Wang, G.~Liu, C.~Liu, and Z.~Wang,
	``A pediatric ecg database with disease diagnosis covering 11643 children,''
	\emph{Scientific Data}, vol.~12, no.~1, p. 867, 2025.
	
	\bibitem{he2016deep}
	K.~He, X.~Zhang, S.~Ren, and J.~Sun, ``Deep residual learning for image
	recognition,'' in \emph{Proceedings of the IEEE conference on computer vision
		and pattern recognition}, 2016, pp. 770--778.
	
	\bibitem{hochreiter1997long}
	S.~Hochreiter and J.~Schmidhuber, ``Long short-term memory,'' \emph{Neural
		computation}, vol.~9, no.~8, pp. 1735--1780, 1997.
	
	\bibitem{vaswani2017attention}
	A.~Vaswani, N.~Shazeer, N.~Parmar, J.~Uszkoreit, L.~Jones, A.~N. Gomez,
	{\L}.~Kaiser, and I.~Polosukhin, ``Attention is all you need,''
	\emph{Advances in neural information processing systems}, vol.~30, 2017.
	
	\bibitem{dao2024transformers}
	T.~Dao and A.~Gu, ``Transformers are ssms: Generalized models and efficient
	algorithms through structured state space duality,'' \emph{arXiv preprint
		arXiv:2405.21060}, 2024.
	
	\bibitem{moreno2024ecg}
	P.~A. Moreno-S{\'a}nchez, G.~Garc{\'\i}a-Isla, V.~D. Corino, A.~Vehkaoja,
	K.~Brukamp, M.~Van~Gils, and L.~Mainardi, ``Ecg-based data-driven solutions
	for diagnosis and prognosis of cardiovascular diseases: A systematic
	review,'' \emph{Computers in Biology and Medicine}, vol. 172, p. 108235,
	2024.
	
	\bibitem{jahmunah2021automated}
	V.~Jahmunah, E.~Y.~K. Ng, T.~R. San, and U.~R. Acharya, ``Automated detection
	of coronary artery disease, myocardial infarction and congestive heart
	failure using gaborcnn model with ecg signals,'' \emph{Computers in biology
		and medicine}, vol. 134, p. 104457, 2021.
	
	\bibitem{alamatsaz2024lightweight}
	N.~Alamatsaz, L.~Tabatabaei, M.~Yazdchi, H.~Payan, N.~Alamatsaz, and F.~Nasimi,
	``A lightweight hybrid cnn-lstm explainable model for ecg-based arrhythmia
	detection,'' \emph{Biomedical Signal Processing and Control}, vol.~90, p.
	105884, 2024.
	
	\bibitem{lin2024ventricular}
	Q.~Lin, D.~Ogli{\'c}, M.~J. Curtis, H.-K. Lam, and Z.~Cvetkovi{\'c},
	``Ventricular arrhythmia classification using similarity maps and
	hierarchical multi-stream deep learning,'' \emph{IEEE Transactions on
		Biomedical Engineering}, 2024.
	
	\bibitem{chen2025advancing}
	Y.~Chen, Z.~Huang, and Z.~Feng, ``Advancing few-shot pediatric arrhythmia
	classification with a novel contrastive loss and multimodal learning,''
	\emph{arXiv preprint arXiv:2509.19315}, 2025.
	
	\bibitem{loshchilov2017decoupled}
	I.~Loshchilov and F.~Hutter, ``Decoupled weight decay regularization,''
	\emph{arXiv preprint arXiv:1711.05101}, 2017.
	
\end{thebibliography}

\end{document}